\RequirePackage{amsmath}
\documentclass[12pt]{iopart}

\usepackage{bm}
\usepackage{graphicx}
\usepackage{cite}
\usepackage{color}
\usepackage{amssymb}
\usepackage{bbold}

\DeclareMathOperator{\sgn}{sgn}

\begin{document}

\title[Electric control of the bandgap in quantum wells with band-inverted junctions]%
{Electric control of the bandgap in quantum wells with band-inverted junctions}

\author{A. D\'{i}az-Fern\'{a}ndez$^{1,2}$, Leonor Chico$^{3}$ and F. Dom\'{i}nguez-Adame$^{1,2}$}

\address{$^{1}$\ GISC, Departamento de F\'{\i}sica de Materiales, Universidad Complutense, E--28040 Madrid, Spain}

\address{$^{2}$\ Department of Physics, University of Warwick, Coventry, CV4 7AL, 
United Kingdom}

\address{$^{3}$\ Instituto de Ciencia de Materiales de Madrid, Consejo Superior de Investigaciones Cient\'{\i}ficas, C/ Sor Juana In\'{e}s de la Cruz 3, E--28049 Madrid, Spain}

\ead{alvaro.diaz@ucm.es}

\date{\today}

\begin{abstract}

In IV-VI semiconductor heterojunctions with band-inversion, such as those made of Pb$_{1-x}$Sn$_{x}$Te or Pb$_{1-x}$Sn$_{x}$Se, interface states are properly described by a two-band model, predicting the appearance of a Dirac cone in single junctions. However, in quantum wells the interface dispersion is quadratic in momentum and the energy spectrum presents a gap. We show that the interface gap shrinks under an electric field parallel to the growth direction. Therefore, the interface gap can be dynamically tuned in experiments on double-gated quantum wells based on band-inverted compounds.

\end{abstract}


\vspace{2pc}

\submitto{\JPCM}

\maketitle

\section{Introduction}

Narrow-gap semiconductors having conduction and valence bands of opposite parity, like Pb$_{1-x}$Sn$_{x}$Te and Pb$_{1-x}$Sn$_{x}$Se, or orbital character, as Hg$_{1-x}$Cd$_x$Te, may undergo band inversion under compositional variation. It is feasible to grow heterojunctions (for instance, PbTe/Pb$_{1-x}$Sn$_{x}$Te with $x>0.36$,  PbSe/Pb$_{1-x}$Sn$_{x}$Se with $x>0.14$ or HgTe/CdTe) where the fundamental gap, defined as the difference between the band-edge energy of the bands with a given orbital character or parity, has opposite sign on each semiconductor. Such band-inverted junctions received much attention because a treatment of the simplest two-band approximation predicted the occurrence of midgap subbands of electron-like and hole-like interface states~\cite{Volkov85,Korenman87,Agassi88,Pankratov90}. These midgap subbands were found to be gapless with linear dispersion, resembling a two-dimensional Dirac cone. 

Band inversion is an essential ingredient in topological insulators. Consequently, since the advent of the topological band theory there is a renewed interest in band-inverted junctions made of II-VI and IV-VI compound semiconductors \cite{Hasan2010,Bansil2016}. In 2006, Bernevig \emph{et al.}~\cite{Bernevig06} studied theoretically the confined states in HgTe/CdTe quantum wells. HgTe is an inverted-band material and CdTe is a normal-band one, so interface states are expected at each junction. Additionally, varying the thickness of the HgTe layer leads to an inversion of the quantum-well hole-like and electron-like subbands. They predicted the occurrence of a topological phase transition at a critical value of the thickness of the quantum well, giving rise to the concept of topological insulator~\cite{Shen12,Bernevig13}. Such prediction was experimentally confirmed shortly afterwards \cite{Konig07}.

In this paper we study interface states in a band-inverted quantum well of IV-VI semiconductors using a two-band model when an external electric field is applied along the growth direction. We do not consider the quantum-well states confined in the middle layer; rather, we concentrate in the behavior of the interface states, which present a linear dispersion relation in single heterojunctions.
Our main results can be summarized as follows: (i)~In contrast to the single junction, the dispersion relation is quadratic in the interface momentum and a gap opens. (ii)~Gap opening arises from the coupling between the Dirac cones of the two interfaces due to the finite width of the quantum well. Most importantly, (iii)~the interface gap shrinks upon increasing the electric field, so that its magnitude can be substantially modified in experiments. Thus, the electric field can be considered as an external way to modify the coupling of the interface states.

\section{Theoretical model}   \label{sec:model}

The two-band model is a reliable approach to obtain the electron states near the band edges in narrow-gap IV-VI semiconductors, for which the coupling to other bands is negligible \cite{Melngailis72,Burkhard79,Agassi88,Assaf16}. It can even be applied to certain III-V semiconductors if such band coupling is small \cite{Zawadzki11}.  The electron wave function is written as a sum of products of band-edge Bloch functions with slowly varying envelope functions. The corresponding envelope function ${\bm\chi}({\bm r})$ is a four-component column vector composed by the two-component spinors ${\bm\chi}_{+}({\bm r})$ and ${\bm\chi}_{-}({\bm r})$ belonging to the two bands. Electron states near the band edges are determined from the Dirac-like equation $\mathcal{H}{\bm\chi}({\bm r})=E{\bm\chi}({\bm r})$ with~\cite{Agassi88,Pankratov90} 
\begin{equation}
\mathcal{H}=v_{\bot}{\bm\alpha}_{\bot}\cdot{\bm p}_{\bot}+v_z\alpha_z p_z
+\frac{1}{2}\,E_{\mathrm{G}}(z)\beta+V_{\mathrm C}(z)\ ,
\label{eq:01}
\end{equation}
where the $Z$ axis is parallel to the growth direction $[111]$. It is understood that the subscript $\bot$ in a vector indicates the nullification of its $z$-component. $E_{\mathrm{G}}(z)$ denotes the position-dependent gap and $V_{\mathrm C}(z)$ gives the position of the gap center. ${\bm\alpha}=(\alpha_x,\alpha_y,\alpha_z)$ and $\beta$ denote the usual $4\times 4$ Dirac matrices
\begin{equation*}
\alpha_i=\begin{pmatrix}
\mathbb{0}_2 & \sigma_i \\
\sigma_i & \mathbb{0}_2
\end{pmatrix} \ ,
\quad
\beta=\begin{pmatrix}
\mathbb{1}_2 & \mathbb{0}_2 \\
\mathbb{0}_2 & -\mathbb{1}_2
\end{pmatrix} \ ,
\quad 
i=x,y,z\ ,
\end{equation*}
$\sigma_i$ being the Pauli matrices, and $\mathbb{1}_n$ and $\mathbb{0}_n$ are the $n\times n$ identity and null matrices, respectively. Here $v_{\bot}$ and $v_z$ are interband matrix elements having dimensions of velocity. Although they may be different in general, we assume isotropic semiconductors and define $v=v_{\bot}=v_z$ hereafter.

In order to keep the algebra as simple as possible, we restrict ourselves to the symmetric situation with same-sized and aligned gaps [$V_\mathrm{C}(z)=0$]. This is not a serious limitation but the calculations are largely simplified. Thus, a single and abrupt interface presents the following profile for the magnitude of the gap 
\begin{equation}
E_{\mathrm{G}}(z)=2\Delta\sgn(z)\ ,
\label{eq:02}
\end{equation}
where $\sgn(z)=\theta(z)-\theta(-z)$ is the sign function and $\theta(z)$ is the Heaviside step function. The envelope function decays exponentially with distance at each side ${\bm\chi}({\bm r})={\bm\chi}(z) \exp\big(i{\bm r}_{\bot}\cdot{\bm k}_{\bot}\big)$ with~\cite{Adame94}
\begin{subequations}
\begin{equation}
{\bm\chi}(z)\sim\exp\left(-\,\frac{|z|}{d}\right)\ ,
\qquad
d=\frac{\hbar v}{\Delta}\ ,
\label{eq:03a}
\end{equation}
and the dispersion is linear in the interface momentum (see, e.g., Ref.~\cite{Pankratov90})
\begin{equation}
E({\bm k}_{\bot})=\pm\hbar v|{\bm k}_{\bot}|\ .
\label{eq:03b}
\end{equation}
\label{eq:03}
\end{subequations}

\section{Quantum well with band inversion}   \label{sec:unbiased}

\begin{figure}
\centerline{\includegraphics[width=0.5\columnwidth]{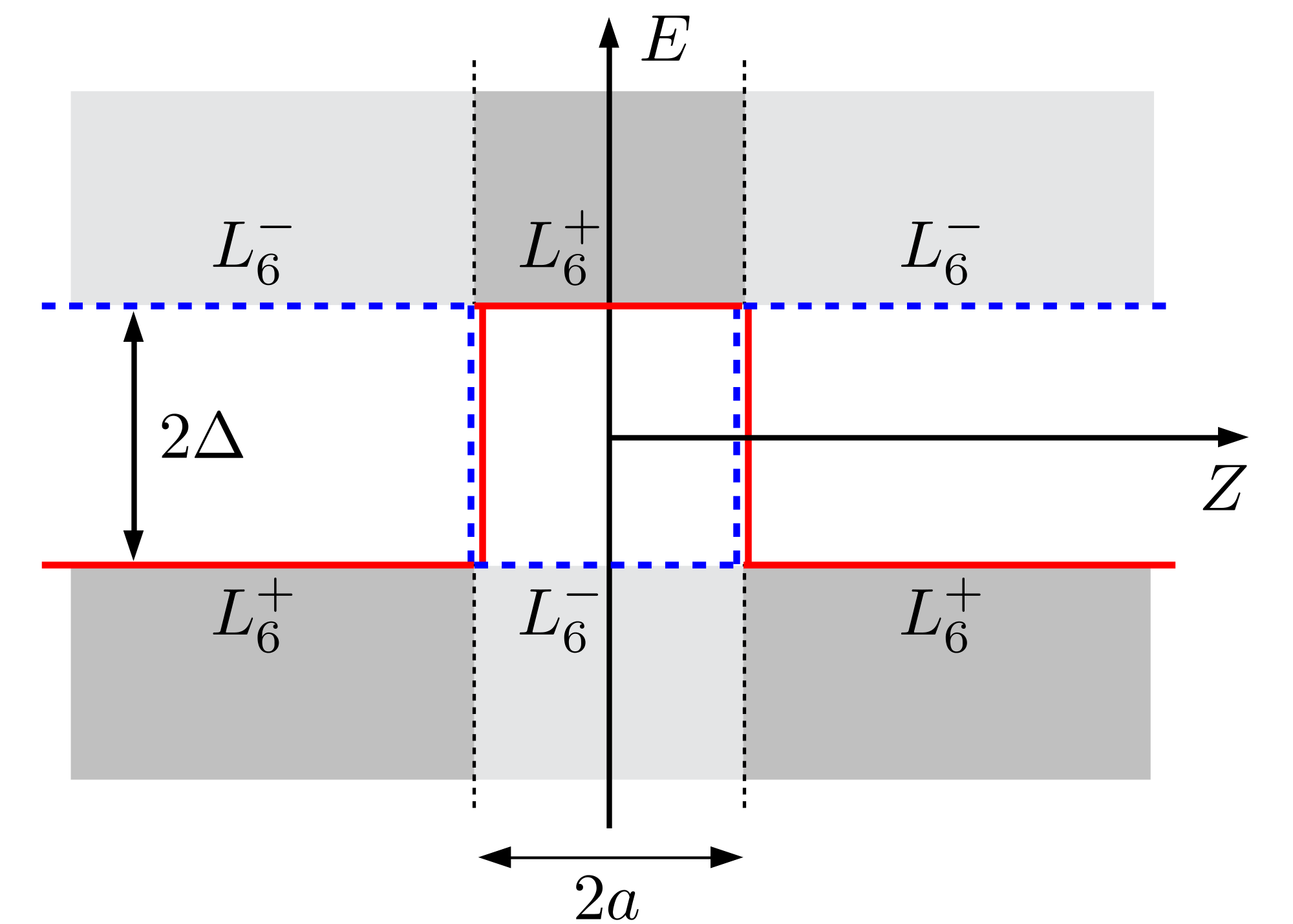}}
\caption{$L_6^{+}$ and $L_6^{-}$ band-edge profile of two band-inverted junctions with aligned and same-sized gaps, located at the $XY$ plane. The distance between the junctions is $2a$ and the magnitude of the gap is $2\Delta$.}
\label{fig0}
\end{figure}

For completeness, in this section we present and discuss the salient features of a quantum well of width $2a$ with band-inversion in the absence of an applied electric field. We introduce an alternative derivation of the interface states, although the final results agree with those obtained in Ref.~\cite{Korenman87}. Assuming that the interface states spread over distances much larger than the interface region, we can consider an abrupt profile for the two band-inverted junctions forming the quantum well. Therefore, the gap profile is now given by
\begin{equation}
E_{\mathrm{G}}(z)=2\Delta\Big[1-2\theta(z+a)+2\theta(z-a)\Big]\ ,
\label{eq:04}
\end{equation}
as depicted in Figure~\ref{fig0}, where we have taken the inverted semiconductor embedded in the non-inverted one. 

Electronic states of the Hamiltonian~(\ref{eq:01}) can be addressed with the aid of the Feynman-Gell-Mann \textit{ansatz} as follows~\cite{Feynman58}
\begin{equation}
{\bm\chi}(z)\!=\!\Big[\hbar v \Big(\!
-i\alpha_z \frac{d\phantom{z}}{dz}
+{\bm\alpha}_{\bot}\cdot{\bm k}_{\bot}\!\Big)
+\frac{1}{2}\,E_{\mathrm{G}}(z)\beta+E\Big]{\bm\psi}(z)\ .
\label{eq:05}
\end{equation}
Defining the following dimensionless quantities ${\bm\kappa}={\bm k}_{\bot}d$, $\xi=z/d$, $\xi_0=a/d$, $\varepsilon=E/\Delta$, and by applying the Hamiltonian~(\ref{eq:01}) to~(\ref{eq:05}), we obtain
\begin{subequations}
\begin{equation}
\left[-\frac{d^2}{d\xi^2}+U(\xi)+\lambda^2\right]{\bm \psi}(\xi)=0\ ,
\label{eq:06a}
\end{equation}
with
\begin{equation}
U(\xi)=2i\beta\alpha_z \Big[\delta(\xi-\xi_0)-\delta(\xi+\xi_0)\Big]\ ,
\label{eq:06b}
\end{equation}
and
\begin{equation}
\lambda^2=\kappa^2+1-\varepsilon^2\ .
\label{eq:06c}
\end{equation}
\label{eq:06}
\end{subequations}
We have used the anticommutation relations of the Dirac matrices and $d\theta(\xi)/d\xi=2\delta(\xi)$. 

We can find exactly the electron energy by means of the Green's function approach. To this end, we can treat the term $U(\xi)$ in Eq.~(\ref{eq:06a}) as a perturbation. The retarded Green's function for the unperturbed problem satisfies
\begin{subequations}
\begin{equation}
\left[-\frac{\partial^2}{\partial \xi^2}+\lambda^{2}\right]
\mathcal{G}_{0}^{+}(\xi,\xi^{\prime};\varepsilon)=\delta(\xi-\xi^{\prime})\mathbb{1}_4\ ,
\label{eq:07a}
\end{equation}
which can be factorized as $\mathcal{G}_{0}^{+}(\xi,\xi^{\prime};\varepsilon)=G_{0}^{+}(\xi,\xi^{\prime};\varepsilon)\mathbb{1}_4$ and it is understood that $\mathrm{Im}(\lambda^2)<0$. Since we are interested in midgap states, we consider $\mathrm{Re}(\lambda^2)>0$. The Green's function for the free particle problem is known to be~\cite{Economou06}
\begin{equation}
G_{0}^{+}(\xi,\xi^{\prime};\varepsilon)=\frac{1}{2\lambda}\exp\left(-\lambda|\xi-\xi^{\prime}|\right) \ .
\label{eq:07b}
\end{equation}
\label{eq:07}
\end{subequations}
We can now apply Dyson's equation to obtain the complete Green's function $\mathcal{G}^{+}(\xi,\xi^{\prime};\varepsilon)$ associated to Eq.~(\ref{eq:06a}) as follows
\begin{eqnarray}
\mathcal{G}^{+}(\xi,\xi^{\prime};\varepsilon) & = & 
\mathcal{G}_{0}^{+}(\xi,\xi^{\prime};\varepsilon) +
\int d\xi^{\prime\prime} \mathcal{G}_{0}^{+}(\xi,\xi^{\prime\prime};\varepsilon)
\nonumber \\
& \times &
U(\xi^{\prime\prime})\mathcal{G}^{+}(\xi^{\prime\prime},\xi^{\prime};\varepsilon)
\ .
\label{eq:08}
\end{eqnarray}

Dyson's equation~(\ref{eq:08}) can be exactly solved due to the simple expression of the potential term~(\ref{eq:06b}). The retarded Green's function $\mathcal{G}^{+}(\xi,\xi^{\prime};\varepsilon)$ is analytic in the lower half plane $\mathrm{Im}(\lambda^2)<0$. Thus, it may have simple poles when it is analytically continued to the upper half plane. After some straightforward algebra, the poles are obtained from the scalar Green's function by solving the following equation
\begin{eqnarray}
\Big[
1 &-& 4G_{0}^{+}(\xi_0,\xi_0;\varepsilon)G_{0}^{+}(-\xi_0,-\xi_0;\varepsilon)
\nonumber\\
&+&4G_{0}^{+}(\xi_0,-\xi_0;\varepsilon)G_{0}^{+}(-\xi_0,\xi_0;\varepsilon)\Big]^2
\nonumber\\
& = &4 \Big[
G_{0}^{+}(\xi_0,\xi_0;\varepsilon)-G_{0}^{+}(-\xi_0,-\xi_0;\varepsilon)\Big]^2\ .
\label{eq:09}
\end{eqnarray}
Recalling Eq.~(\ref{eq:07b}), we get $\lambda^{2}-1+\exp(-4\lambda \xi_0)=0$. For not too narrow quantum wells and reverting the change of variables we finally are arrive at
\begin{equation}
E({\bm k}_{\bot})=\pm\sqrt{\hbar^2 v^2 k_{\bot}^2+\Delta^2 \exp\left(-4a/d\right)} \ .
\label{eq:10}
\end{equation}
The dispersion is no longer linear and an interface gap of magnitude $2\Delta_{w0}$ opens, where
\begin{equation}
\Delta_{w0}=\Delta \exp\left(-2\,\frac{a}{d}\right)\ .
\label{eq:11}
\end{equation}
The subscript $0$ refers to the absence of applied electric field. The gap is due to the coupling of the two interface states 
arising at the well boundaries. This finite-size effect turns the interface Dirac fermions massive \cite{Zhou2008,Lu2010}.  

\section{Quantum well under bias}   \label{sec:biased}

Now we turn to the interface states of a quantum well with band-inversion subjected to a uniform electric field ${\bm F}=-F\,\widehat{\bm z}$, following the approach introduced in Ref.~\cite{Diaz-Fernandez17}. The Dirac equation then reads $\big(\mathcal{H}-eFz\big){\bm\chi}({\bm r})=E{\bm\chi}({\bm r})$, where $\mathcal{H}$ is given in~(\ref{eq:01}). The Feynman-Gell-Mann \emph{ansatz}~(\ref{eq:05}) with the replacement $E\to E+eFz$ renders the Dirac-like equation into a Schr\"{o}dinger-like equation
\begin{equation}
\left[-\frac{d^2}{d\xi^2}+U(\xi)-f^2\xi^2-if\alpha_z-2\varepsilon f\xi
+\lambda^2\right]{\bm \psi}(\xi)=0\ .
\label{eq:12}
\end{equation}
where $f=F/F_{\mathrm{C}}$ and $F_{\mathrm{C}}=\Delta/ed=\Delta^2/e\hbar v$. The term $-f^2\xi^2$ is negligible under the assumption that $F < F_{\mathrm{C}}$ because the envelope function is vanishingly small if $\xi > 1$. Note that this is the usual regime in experiments since typical values for IV-VI compounds are $\Delta=75\,$meV and $d=4.5\,$nm~\cite{Korenman87}, yielding $F_\mathrm{C}=170\,$kV/cm. Regarding the constant matrix term $-if\alpha_z$, it is easily diagonalized by a unitary transformation. Nevertheless, we have checked that it has a small impact on the final results even at moderate fields~\cite{Diaz-Fernandez17}. Thus, we omit those two terms in what follows.

We can regard again the term $U(\xi)$ in~(\ref{eq:12}) as a perturbation and seek for the retarded Green's function of the unperturbed problem $\mathcal{G}_{0}^{+}(\xi,\xi^{\prime};\varepsilon)=G_{0}^{+}(\xi,\xi^{\prime};\varepsilon)\mathbb{1}_4$, where the scalar Green's function obeys the following equation
\begin{equation}
\left[-\frac{\partial^2}{\partial\xi^2}-2\varepsilon f\xi
+\lambda^2\right]G_{0}^{+}(\xi,\xi^{\prime};\varepsilon)=
\delta(\xi-\xi^{\prime})\ .
\label{eq:13}
\end{equation}
Equation~(\ref{eq:13}) is analogous to the problem of a non-relativistic particle in a tilted potential solved in Refs.~\cite{Ludviksson87,Jung09}. Let us define
\begin{equation}
\mu=(2|\varepsilon| f)^{1/3}\ ,
\quad
p(\xi)=-s_{\varepsilon}\mu\,\xi+\frac{\lambda^2}{\mu^2}\ .
\label{eq:14}
\end{equation}
with the shorthand notation $s_{\varepsilon}=\mathrm{sgn}\left[\mathrm{Re}(\varepsilon)\right]$. In terms of these parameters the retarded Green's function is written as
\begin{eqnarray}
G_{0}^{+}(\xi,\xi^{\prime};\varepsilon)&=&-\frac{\pi s_{\varepsilon}}{\mu}\Big\{
\theta\left[(\xi^{\prime}-\xi)s_{\varepsilon}\right]
\mathrm{Ai}\left(p(\xi)\right)
\mathrm{Ci}^{+}\left(p(\xi^{\prime})\right)
\nonumber\\
&+&
\theta\left[(\xi-\xi^{\prime})s_{\varepsilon}\right]\mathrm{Ai}\left(p(\xi^{\prime})\right)
\mathrm{Ci}^{+}\left(p(\xi)\right)\Big\}\ ,
\label{eq:15}
\end{eqnarray}
where $\mathrm{Ci}^{+}(z)=\mathrm{Bi}(z)+i\mathrm{Ai}(z)$, $\mathrm{Ai}(z)$ and $\mathrm{Bi}(z)$ being the Airy functions~\cite{Abramowitz72}. It is worth mentioning that $G_{0}^{+}(\xi,\xi^{\prime};\varepsilon)$ is continuous on the line $\xi=\xi^{\prime}$.

Once the retarded Green's function for the unperturbed problem is known, we can get the energy levels from~(\ref{eq:09}). In the presence of the electric field, poles are of the form $E({\bm \kappa}_{\bot})-i\Gamma({\bm \kappa}_{\bot})/2$ and correspond to resonant states. Therefore, electrons can tunnel into the continuum and escape from the quantum well. This is a common feature in the quantum-confined Stark effect~\cite{Miller84}. Nonetheless, the level width is exponentially small in the low-field regime, namely, tunneling is only important at very high fields~\cite{Diaz-Fernandez17}. Thus, we omit the imaginary part hereafter.

\section{Low-field limit}   \label{sec:low-field}

We can simplify~(\ref{eq:15}) in the low-field regime $F<F_\mathrm{C}$ by noticing that $|\lambda(\varepsilon,\kappa)|\gg \mu(\varepsilon)$. In this limiting case we approximate the Airy functions to their asymptotic expansions for large argument~\cite{Abramowitz72}. In this regime we take $\mathrm{Ci}^{+}(z)\simeq\mathrm{Bi}(z)$ and 
\begin{equation}
\mathrm{Ai}(z)  \simeq  \frac{1}{2\sqrt{\pi}}\,
\frac{e^{-\phi}}{z^{1/4}}\,L(-\phi) \ , 
\quad
\mathrm{Bi}(z)  \simeq  \frac{1}{\sqrt{\pi}}\,\frac{e^{\phi}}{z^{1/4}}L(\phi)\ ,
\label{eq:16}
\end{equation}
with $\phi=(2/3)z^{3/2}$ and $L(\phi)=1+\sum_{\ell=1}^{\infty}u_{\ell}\phi^{-\ell}$, where $u_{\ell}=\Gamma(3\ell + 1/2)/54^{\ell}\, \ell!\, \Gamma(\ell + 1/2)$, $\Gamma(z)$ being the $\Gamma$ function. We can now obtain an expression to the lowest order in the field as follows
\begin{eqnarray}
G_{0}^{+}(\pm \xi_0,\pm \xi_0;\varepsilon)&=&-\frac{1}{2\lambda}
\left(1\pm\frac{|\varepsilon|\xi_0}{\lambda^2}f\right)\ ,
\nonumber \\
G_{0}^{+}(\pm \xi_0,\mp \xi_0;\varepsilon)&=&-\frac{1}{2\lambda}\,
\exp\left(-2\lambda \xi_0\right)\ .
\label{eq:17}
\end{eqnarray}
Finally, inserting~(\ref{eq:17}) into~(\ref{eq:09}) yields an approximate expression to obtain the energy of the interface states in the quantum well
\begin{equation}
\varepsilon^2=\kappa^2+\exp\left(-4\lambda \xi_0\right)
-\frac{2|\varepsilon|\xi_0}{\lambda}\,f.
\label{eq:18}
\end{equation}

\begin{figure}[ht]
\begin{center}
\includegraphics[width=0.5\linewidth]{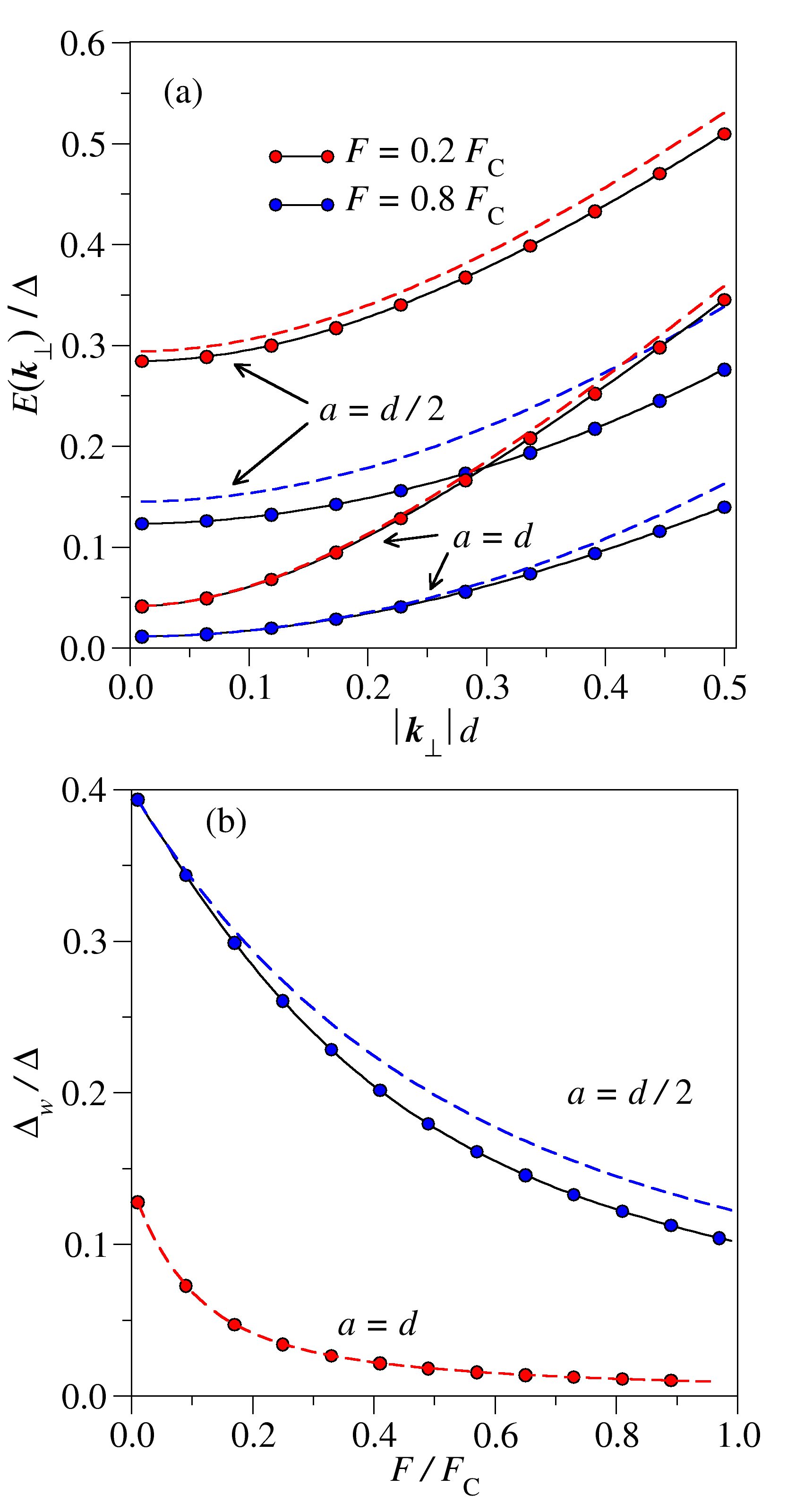}
\end{center}
\caption{(Color online) (a)~Energy in units of $\Delta=E_G/2$ as a function of the 
in-plane 
momentum for two values of the applied field ($F=0.2F_{\mathrm{C}}$ and $F=0.8F_{\mathrm{C}}$) and two widths of the quantum well ($a=d/2$ and $a=d$). Mirror images are obtained for negative energies. (b)~Interface gap $2\Delta_{w}$ in units of the fundamental gap $2\Delta$ as a function of the electric field for two different widths. Dashed lines show the approximate solution given in~(\ref{eq:18}).}
\label{fig1}
\end{figure}

In order to verify the accuracy of the result, we numerically tested~(\ref{eq:18}) from the numerical solution of Eq.~(\ref{eq:09}) using the exact Green's function. Taking $s_\varepsilon=1$ for concreteness, from~(\ref{eq:15}) we get
\begin{eqnarray}
G_{0}^{+}(\pm\xi_0,\pm\xi_0;\varepsilon)
&=&-\frac{\pi}{\mu}\,\mathrm{Ai}(z_{\pm})\mathrm{Ci}^{+}(z_{\pm})\ ,
\nonumber \\
G_{0}^{+}(\pm\xi_0,\mp\xi_0;\varepsilon)
&=&-\frac{\pi}{\mu}\,\mathrm{Ai}(z_{-})\mathrm{Ci}^{+}(z_{+})\ ,
\label{eq:19}
\end{eqnarray}
with $z_{\pm}=\lambda^2/\mu^2\mp\mu\xi_0$. Figure~\ref{fig1}(a) shows the dispersion relation for two values of the applied field ($F=0.2F_{\mathrm{C}}$ and $F=0.8F_{\mathrm{C}}$) and two widths of the quantum well ($a=d/2$ and $a=d$). 
Dashed lines show the approximate low-field limit~(\ref{eq:18}). We conclude that the analytical result fits the numerics quite well except at high field ($F=0.8F_{\mathrm{C}}$) and small width ($a=d/2$), as expected. From the dispersion relation we can obtain the gap of the interface states $\Delta_w$ as the difference of the positive and negative energy solutions at ${|{\bm k}_{\bot}|}\to 0$. Figure~\ref{fig1}(b) shows that this gap shrinks upon increasing the electric field. Therefore, we come to the conclusion that the gap can be controlled to a large extent by the field.

It is worth mentioning that Eq.~(\ref{eq:18}) can be further simplified when the quantum well is not too narrow. In this case we can take $\lambda\simeq 1$. Reverting the change of variables we get
\begin{equation}
E({\bm k}_{\bot})=\pm \Big[\sqrt{(eFa)^2+\hbar^2 v^2 k_{\bot}^2
+\Delta_{w0}^2}
-eFa\Big]\ ,
\label{eq:20}
\end{equation}
where $\Delta_{w0}$ is given by~(\ref{eq:11}). Notice that turning off the field we recover Eq.~(\ref{eq:10}). The gap of the 
interface states 
 is then approximately given as $2\Delta_w$ with
\begin{equation}
\Delta_w = \sqrt{(eFa)^2+\Delta_{w0}^2}-eFa\ .
\label{eq:21}
\end{equation}
Equation~(\ref{eq:21}) is very remarkable and it is our main result. Although being approximate, we have found that it is very accurate unless the field is high and the quantum well is narrow. It implies that applying an electric field perpendicular to the junction, the interface gap diminishes. Notice that there exist two different regimes. At low field, i.~e., $eFa<\Delta_{w0}$, the gap decreases linearly as $\Delta_w\simeq \Delta_{w0}-eFa$. On the contrary, at high field the gap vanishes according to the power law $\Delta_w\simeq \Delta_{w0}^2/2eFa$. Since the gap is a consequence of the hybridization of the interface states, the electric field can be viewed as an external means to control the coupling of these bands.

\section{Conclusions}

In this work we have studied band-inverted quantum wells subjected to an electric field applied along the growth direction. We used a spinful two-band model that is equivalent to the Dirac model for relativistic electrons. The mass term is half the bandgap and changes its sign across the junction. In the case of a single band-inverted junction, the envelope function of the interface states is exponentially localized in the growth direction with decay length $d=\hbar v/\Delta$. The corresponding interface dispersion is linear, as given by~(\ref{eq:03b}), and is commonly called a Dirac cone. A second junction at a distance $2a$ not large compared to $d$ yields the splitting of the Dirac cones into two massive subbands and an interface gap opens, as expressed by Eq.~(\ref{eq:11}). Therefore, finite-size effects give mass to the Dirac fermions, transforming their linear dispersion into a parabola at small wave vectors. Remarkably, although the interface gap never closes, it can be dramatically reduced by the electric field. Under certain reasonable assumptions we have found a simple expression for the interface gap as a function of the field, as shown in~(\ref{eq:21}). This expression predicts a linear reduction of the gap if the electric field is smaller than $F_{\text{C}}\exp(-2a/d)d/a$, while it decays as a power law at higher fields. 

\ack

A.~D.-F.\ and F.~D-A.\ thank the Theoretical Physics Group of the University of Warwick for their warm hospitality. This work was supported by the Spanish MINECO under grants MAT2013-46308, MAT2016-75955 and FIS2015-64654-P. 

\section*{References}

\bibliography{references}

\bibliographystyle{iopart-num.bst}

\end{document}